\documentclass[reprint,amsmath,amssymb, aps,]{revtex4-2}

\usepackage{graphicx}
\usepackage{dcolumn}
\usepackage{bm}
\usepackage[normalem]{ulem} 
\usepackage{color}
\usepackage{float}
\usepackage{caption}
\usepackage{subcaption}
\DeclareCaptionJustification{justified}{\leftskip=0pt \rightskip=0pt \parfillskip=0pt plus 1fil}
\usepackage{lineno}
\usepackage{epstopdf}
\usepackage{hyperref}

\begin{document}

\preprint{APS/??}

\title{Swelling and evaporation determine surface morphology of grafted hydrogel thin films}

\author{Caroline Kopecz-Muller$^{1,2,3}$}
\author{Cl\'{e}mence Gaunand$^{1,2}$}
\author{Yvette Tran$^{4,5}$}
\author{Matthieu Labousse$^1$}
\author{Elie Raphaël$^1$}
\author{Thomas Salez$^3$}
\author{Finn Box$^{1,2,6}$}
\email{finn.box@manchester.ac.uk}
\author{Joshua D. McGraw$^{1,2}$}
\email{joshua.mcgraw@espci.fr}

\affiliation{$^1$Gulliver, CNRS, ESPCI Paris, Université PSL, 75005 Paris, France}
\affiliation{$^2$Institut Pierre Gilles de Gennes (IPGG), ESPCI Paris, Université PSL, 75005 Paris, France}
\affiliation{$^3$LOMA, CNRS, Univ. Bordeaux, F-33405 Talence, France}
\affiliation{$^4$Sciences et Ing\'{e}nierie de la Mati\`{e}re Molle, CNRS, ESPCI Paris, Universit\'{e} PSL, 75005 Paris, France}
\affiliation{$^5$Sorbonne-Universit\'{e}s, UPMC Université Paris 06, 75005 Paris, France}
\affiliation{$^6$Physics of Fluids \& Soft Matter, Department of Physics \& Astronomy, University of Manchester, Manchester M13 9PL, United Kingdom}

\begin{abstract}

\textbf{Abstract  --}  We experimentally study the formation of surface patterns in grafted hydrogel films of nanometer-to-micrometer thicknesses during imbibition-driven swelling followed by evaporation-driven shrinking. Creases are known to form at the hydrogel surface during swelling; the wavelength of the creasing pattern is proportional to the initial thickness of the hydrogel film with a logarithmic correction that depends on microscopic properties of the hydrogel. We find that, although the characteristic wavelength of the pattern is determined during swelling, the surface morphology can be significantly influenced by evaporation-induced shrinking. We observe that the elastocapillary length based on swollen mechanical properties gives a threshold thickness for a surface pattern formation, and consequently an important change in morphology.

\end{abstract}

\pacs{??}
              
\maketitle

\section*{Introduction}

Smart hydrogels that respond to external stimuli enable the design of adaptive surface coatings \cite{Russell2002,ahn2008stimuli}.
For example, grafted films of temperature-responsive hydrogels~\cite{Li1994a,kim2010dynamic,Yoon2010} have been used in switchable microfluidic valves with rapid, reliable, and repeatable actuation \cite{Beebe2000,Idota2005} and to trap single cells \cite{DEramo2018}. These applications harness volume changes for functionality by exploiting the ability of polymers, in these cases poly(N-isopropylacrylamide) (PNIPAM), to swell in water, by about a factor 4, at room temperature~\cite{Li2015}. The resulting hydrogel exhibits a temperature-controlled transition between a swollen state, showing hydrophilic properties and elastic moduli in the kilopascal range, and a collapsed state, characterized by a hydrophobic behavior and a 100-fold stiffness increase~\cite{winnik1990fluorescence,heskins1968solution,Li2015,haq2017mechanical}. The switchable physico-chemical properties of PNIPAM led to design media for controlled release in drug delivery systems \cite{Peppas2006}. Furthermore, biocompatibility of PNIPAM is exploited by using films as substrates for cellular cultures and protein adsorption, as a bioscaffold in tissue engineering applications~\cite{Niloofar2016}\footnotetext{Corresponding authors : }.

\begin{figure}[b!]
 	\centering
 	\includegraphics[width=0.99 \linewidth]{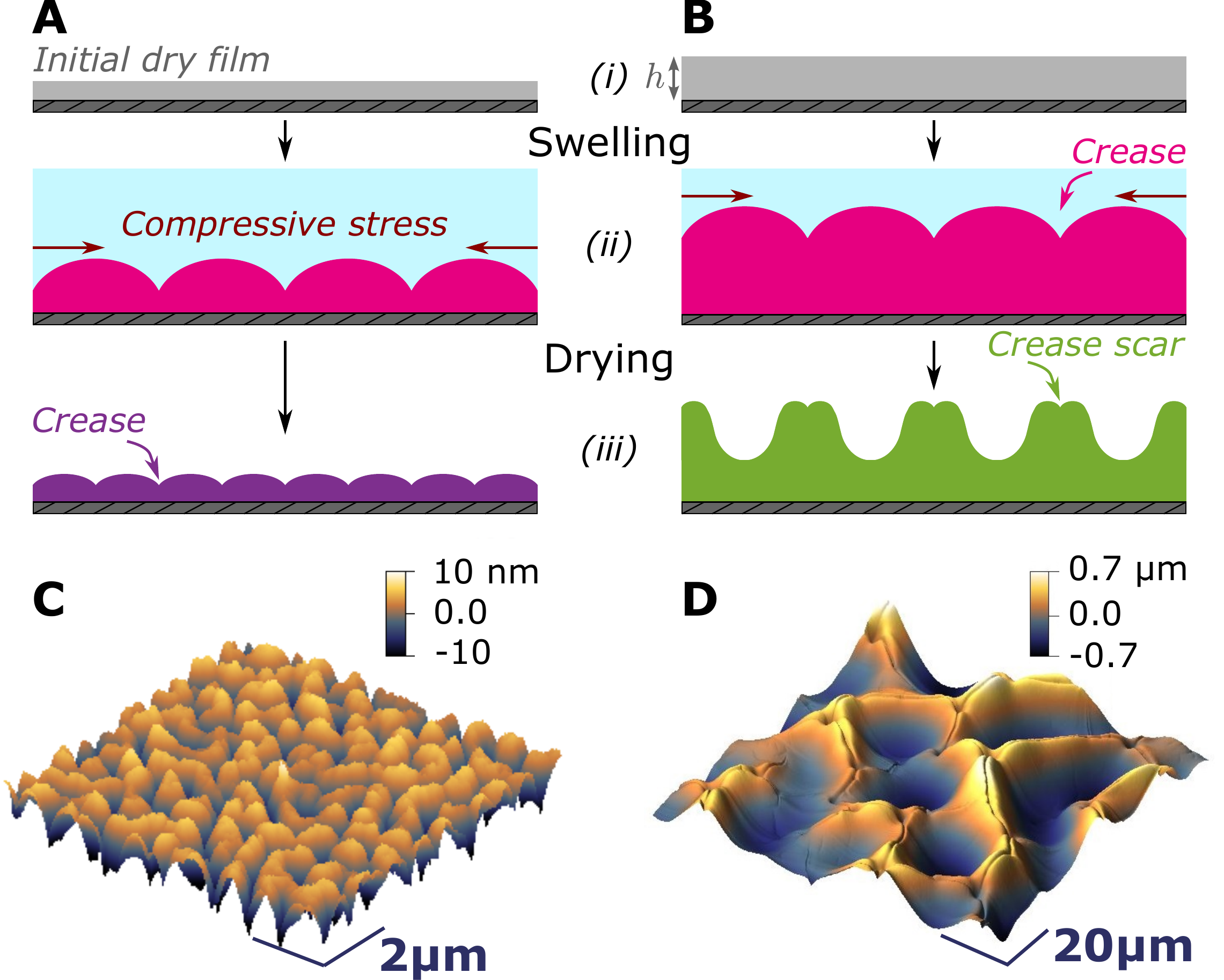}
 	\caption{\small{\textbf{Nanoscale surface patterns in an ultrathin hydrogel films undergoing irreversible swelling/shrinking deformations.} \textit{A) and B) Schematics of observed patterns, for nanometric (A) and micrometric (B) initial thicknesses. C) and D) 3D reconstruction of the surface topography of two PNIPAM films, with initial dry thicknesses of $h=232\,$nm (C) and $h=4.5\,\mu$m (D), that were (\textit{i}) crosslinked and grafted to a rigid substrate, (\textit{ii}) swollen by immersion in water, (\textit{iii}) dried via evaporation and (C and D) imaged using AFM ; colors in C and D correspond to relative surface height (colorbars).}}}
 	\label{fig:3D}
 \end{figure}

Many of these applications require grafting the polymer to a substrate. Such grafting can constrain the swelling of hydrogel films and lead to surface patterning. Indeed, volumetric expansion associated with osmotic pressure is inhibited laterally by substrate-attachement. This geometrical frustration generates an in-plane compressive stress that can destablize the flat surface of the hydrogel \cite{Trujillo2008,Hong2009,Ortiz2010,Chen2014, Ju2022}. This  nonlinear instability \cite{Ciarletta2018, cai2012creasing,Hohlfeld2011} renders the surface topography non-uniform, with sharp `creases' (localized regions of large strain/curvature \cite{Chen2012}) separating smooth peaks \cite{Suematsu1990,Trujillo2008,Li2012}. Such creases are schematized in Figs.~\ref{fig:3D}A~and~B.
 
To date, the majority of the reported surface patterns occurred in wet hydrogel films of micrometric-to-millimetric thickness  \cite{Tanaka1987,Tanaka1992,Li1994,Li1994a,Gerardin2006,Trujillo2008,Durie2015}. Here, we report on a morphological transition on nanometric-to-millimetric gels that occurs upon drying, as illustrated in Figs.~\ref{fig:3D}A and B schematically, and in C and D from atomic force microscopy (AFM).

Surface patterns in soft cross-linked gels were first reported over a century ago~\cite{Sheppard1918} and in swollen rubber in the 60's~\cite{southern1965effect}. A first theoretical prediction for surface instabilities occurring under in-plane compression was made by Biot~\cite{Biot1963}. In this prediction, on the condition that the in-plane strain exceeds a threshold of 0.46, an elastomeric half space deforms in a sinusoidal pattern, referred to as `wrinkles'~\cite{mora2011surface,Biot1963}. Later, Tanaka and co-workers investigated the formation and evolution of creases at the surface of synthetic gels during swelling~\cite{Tanaka1987,Tanaka1992}. Unlike wrinkles that exhibit a sinusoidal shape, creases are characterized by localized regions of large, in-depth, surface curvature.

From the first observations of creases at the surface of gels and elastomers, further theoretical and numerical works~\cite{Hohlfeld2011,hohlfeld2008thesis, Hong2009} as well as experimental works~\cite{Trujillo2008,cai2012creasing} showed that the creasing instability occurs at a critical strain threshold of $\epsilon_\mathrm{c} \sim 0.30 - 0.35$. We note that this latter threshold is lower than Biot's earlier prediction, suggesting that creases appear before wrinkles as strain increases~\cite{Chen2014,Chen2012}.
The creasing instability is triggered by imperfections, experimentally in the form of nanoscale roughness and numerically by introducing defects \cite{Cao2012,Chen2012,wong2010surface,kim2010dynamic}.
 The singular deformation of a crease requires the bulk elastic energy to overcome the barrier of surface tension. The elastocapillary length $l_\mathrm{ec} = \gamma/G$ introduces a lengthscale at which the surface tension $\gamma$ and the shear modulus $G$ are balanced. Creasing thus releases compression, and produces topographic variations that are reminiscent of the structural architecture of the brain \cite{Hohlfeld2011,hohlfeld2008thesis,Tallinen2013}. Swelling-induced creases have been observed in many soft and swollen materials~\cite{Yoon2010,Sheppard1918,Tanaka1987,Tanaka1992,Ortiz2010,Guvendiren2010,southern1965effect,drummond1988surface,Trujillo2008,Ju2022} as well as in compressed elastomers~\cite{cai2012creasing,Chen2012, ghatak2007kink, chen2014bilayer}, that are thicker than the elastocapillary length. In the latter case, destabilization of the surface is reversible as compression is released. In the following paragraph, we highlight the irreversible character of several swelling-drying processes.

 In contrast to swelling matter, drying of complex liquids has been a subject of interest for droplets~\cite{deegan2000pattern,zang2019evaporation,ozawa2005modeling,pauchard2003stable} and films~\cite{de2001instabilities,de2002solvent,routh2013drying,okuzono2006simple}. Drying of solutions found application in the development of the spin-coating technique~\cite{bornside1989spin} or in the study of the drying and shrinking of food products~\cite{fu2019differential}.  Deswelling of hydrogels has received less attention in the literature than swelling, and previous studies have focused on macroscopic systems \cite{Bertrand2016,Etzold2021,Engelsberg2013}.  Although in a drying process the ambient conditions and the initial and final states are precisely reversed compared to swelling, drying is only the reverse process of swelling if it occurs very slowly and if elastic deformations are small with respect to the gel size~\cite{Bertrand2016}.  
 Nonlinearities that arise through large deformation can render swelling and deswelling as asymmetric processes \cite{Bertrand2016,Etzold2021}. These result in non-reciprocal and irreversible deformations in a single swelling-deswelling cycle~\cite{Yoon2010}. 
 
 In particular during swelling, the intake of solvent takes place from the free surface and the swollen hydrogel near the surface thus become more permeable. In contrast, evaporation reduces the fraction of solvent, may even lead to an inhomogeneity of solvent concentration with enhanced polymer concentration at the surface~\cite{deegan2000contact,routh2013drying,deegan2000pattern,hennessy2017minimal}, and thus slower dynamics for the solvent in the near-surface region. The latter phenomenon is reminiscent of the accumulation of solute near the contact line found in drying of thin films of colloidal solutions~\cite{deegan2000contact,routh2013drying,deegan1997capillary,hennessy2017minimal}. Phase changes, such as glass transition may occur~\cite{de2002solvent,pauchard2003stable,zang2019evaporation} in a region localized near the surface of polymer films, leading to the formation of a skin layer or ``crust''~\cite{de2002solvent,okuzono2006simple,pauchard2003buckling,pauchard2003stable}, akin to that seen in drying sessile drops with a pinned contact line~\cite{deegan2000contact,zang2019evaporation,larson2014transport,hu2005analysis,hu2005analysisMarangoni,poulard2007control}. Buckling instabilities, which are characterized by the whole surface between two pinned lines growing and becoming more curved, have been reported for drying polymer films~\cite{pauchard2003mechanical,pauchard2003buckling,pauchard2003stable} and colloidal solutions~\cite{pauchard2003mechanical,zhou2017structure,Vermant_2005}, and non-uniform drying may even generate complex structures on dry films such as rings~\cite{chen2009arched,mcgraw2010plateau,mcgraw2011dynamics} or cracks~\cite{xie2021delamination,routh2013drying}, among other patterns~\cite{deegan2000pattern,larson2014transport}. However, creased patterns on dry films, specifically, have been rarely reported~\cite{Ortiz2010, Saintyves2023}.
 
 \begin{figure*}
 	\centering
 	\includegraphics[width=0.95\linewidth]{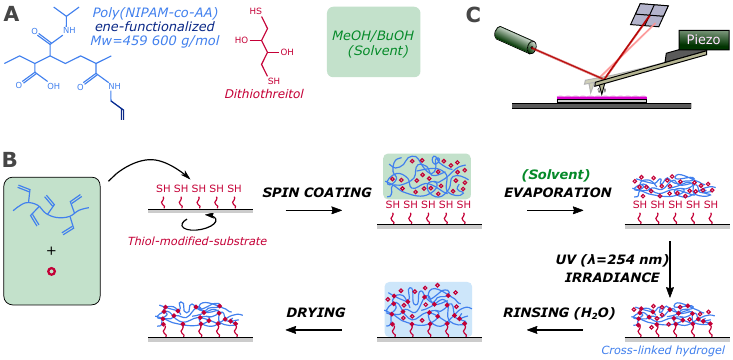}
 	\caption{\small{\textbf{Synthesis by click-chemistry of ultrathin hydrogel films with nanoscale surface patterns.}\textit{ A) P(NIPAM-co-AA) is dissolved in a methanol-butanol mixture (50\,\%/50\,\% in volume) with dithioerythitol, a crosslinker agent. B) Uniform thin films of PNIPAM are achieved through spin-coating and solvent evaporation. The hydrogel films are simultaneously cross-linked and surface-grafted by using thiol-ene click reaction achieved under UV irradiation~\cite{kolb2001click, chollet2016tailoring, benjamin2016multiscale}. Initially flat, dry films are then swollen in water and deswollen via evaporative drying. C) The surface topography of the resulting patterned films is measured using AFM, in both dry and wet conditions. Figure adapted from~\cite{kopecz2024mechanics}, available in Open Access on \url{https://theses.hal.science/}.} }}
 	\label{fig:synthesis}
 \end{figure*}
 
Although characteristic features of the creasing instability can be observed in experiments, the mathematical modeling of the phenomenon is challenging. Intrinsic non-linearity is revealed by the singular shape, then classical linear stability analysis has to be complemented by introducing nucleation and taking into account capillarity in boundary conditions~\cite{Hong2009,wong2010surface,Chen2012,Yoon2010,cai2012creasing}. The latter capillarity involves the effective surface tension between air and the imbibing solvent~\cite{BenAmar2010}. A macroscopic wavelength of instability~\cite{Dervaux2011} for uniaxial swelling of hydrogels under geometric constraints is identified and roughly scales linearly with the thickness. A finer scaling includes a logarithmic correction, comparing the thickness to the dry elastocapillary lengthscale $l_\mathrm{ec}^\text{dry}$, given by the ratio between the effective surface tension between the swollen gel and air and the shear modulus of the solid in the initial dry state $l_\mathrm{ec}^\text{dry} = \gamma/G^\text{dry}$~\cite{Kang2010,Mora2011,Chen2012,Liu2019,vanLimbeek2021}. The model of Dervaux, Ben Amar and Ciarletta \cite{BenAmar2010,Dervaux2011} is based on the hypotheses that volumetric expansion of the gel associated with swelling is limited by the elasticity of the the bulk layer and the surface. Using functional minimization techniques, and imposing a periodic final shape of the surface, they derived the aforementioned scaling with logarithmic correction. This model compares favorably with patterns in swollen films, experimentally performed with an imposed volumetric growth \cite{Tanaka1987,Tanaka1992,Li1994,Li1994a} and an imposed chemical potential \cite{Gerardin2006}.

The article is organized as follows: we first describe the sample fabrication and the AFM-based measurement technique. We then investigate typical AFM images of dry and wet PNIPAM films, showing a transition from a creased pattern to a more sinusoidal morphology with what we refer to as crease scars. We present a quantitative analysis of these results and discriminate different regimes associated with specific morphologies. We show that hydration-dependent elastic properties determine thresholds for pattern formation.  We finally discuss the role evaporation may play, in determining the final surface morphology.\\

\section*{Experiments}
We prepare PNIPAM surface-grafted hydrogel films using the procedure outlined in Fig.~\ref{fig:synthesis} (see supplementary information, SI, Sec.~I.A-D for further details). An in-house synthetized PNIPAM, functionalized with ene-reactive groups (weight-average molar mass $M= 459\,600$\,g/mol, polydispersity index about 2) \cite{Li2015, chollet2016tailoring, benjamin2016multiscale}, is used throughout this study. Uniform films of ene-functionalized PNIPAM are produced by spin-coating from a solution onto the surface of thiolated substrates (silicon wafers or glass substrates). Two orders of magnitude variation in film thickness are achieved by spin-coating at different angular velocities, from 1000 to 4000\,rpm, and by varying the PNIPAM concentration in the initial polymer solution, from 0.5\,\% to 15\,\% (w/w, see SI, Sec.~I.C). The hydrogel coating is cross-linked under deep-UV irradiance (see SI, Sec.~I.D). To remove the excess crosslinker and to induce swelling, samples are immersed in deionized water for 4\,h then in isopropanol (HPLC grade) for 15\,min. Finally, the samples are dried under ambient conditions.

The surface topography of dry and wet films is measured using an Atomic Force Microscope (Nanosurf CoreAFM). The average thicknesses of dry and wet films are also measured by AFM, after a scalpel is used to cut a line across the dry hydrogel film, exposing the substrate. Finally, the Young's modulus is extracted from AFM spectroscopy measurements (see SI, Sec.~III.A).
\\

Several tools were used to rationalize the distinction between brain-like and volcano patterns. In the following, we present these tools and properly define the wavelength $\lambda$ and amplitude $A$.
\begin{figure}[t!]
	\centering
	\includegraphics[width =1\linewidth]{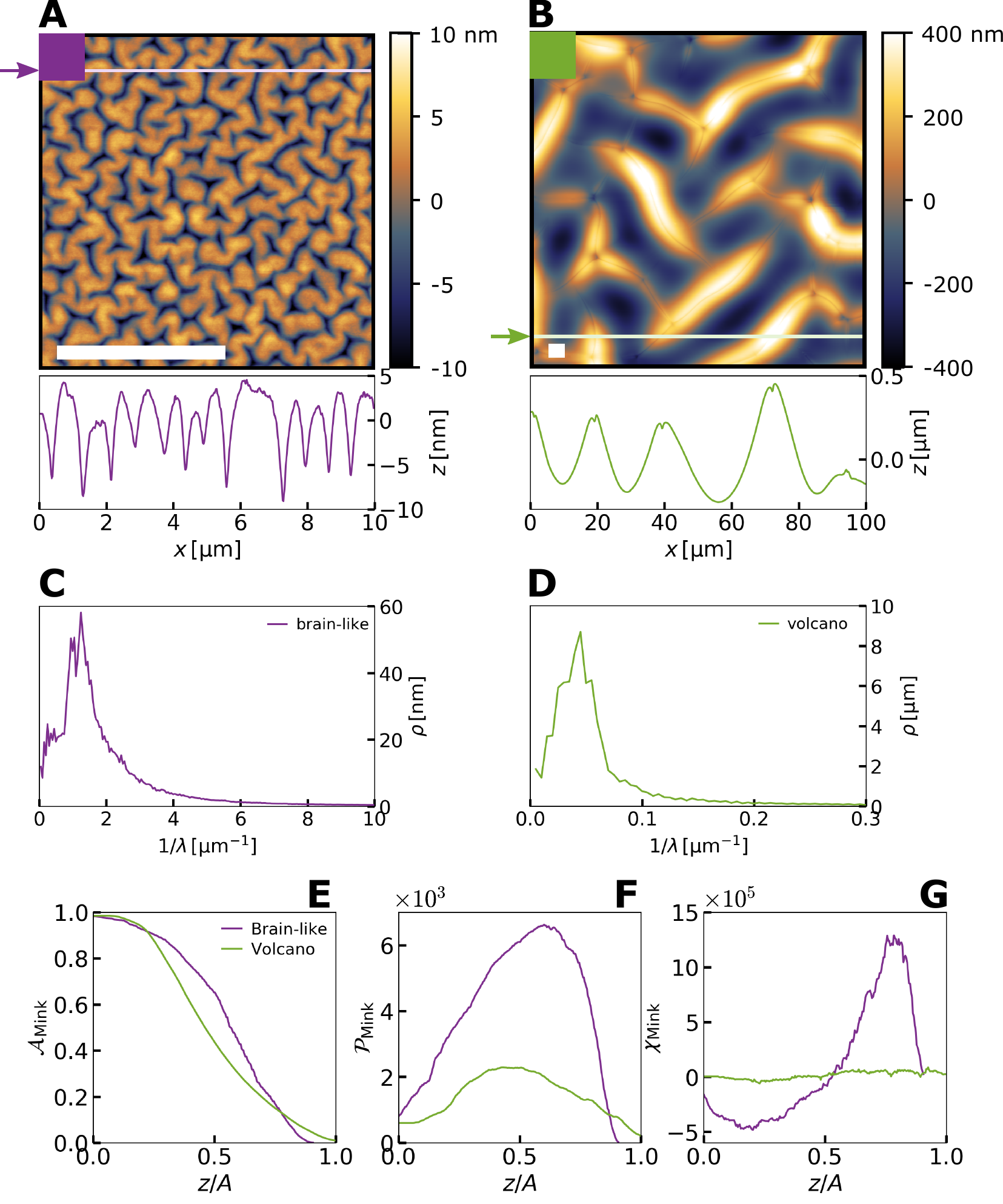}
	\caption{\small{\textbf{Examples of height profiles, Fourier spectra and Minkowski functionals obtained on typical brain-like (left) and volcano (right) pattern images}. \textit{A) and B) Typical AFM images and height profile extracted along the indicated line. The white bar represents $5\,\mathrm{\mu m}$. C) and D) radial averages of 2D Fourier transforms of images showed respectively in panels A) and B) The obtained spectra are showed as a function of the wavenumber $1/\lambda$. Minkowski functionals, calculated from the images showed in panels A) and B). E) Minkowski area $\mathcal{A}_\mathrm{Mink}$, F) perimeter  $\mathcal{P}_\mathrm{Mink}$,  and G) connectivity $\chi_\mathrm{Mink}$ measured as a function of height $z$, normalized by the amplitude of surface topography $A$.}}}
	\label{fig:profiles}
\end{figure}

\paragraph*{Height profiles:~}From an AFM measurement of the surface topography, the height profile along a line is extracted as showed in Figs.~\ref{fig:profiles}A and B, then a wavelength $\lambda$ and an amplitude $A$ are computed. For a given image, between 10 and 20 profiles are extracted, and measurements of wavelength and amplitude are averaged. The resulting error on the measured wavelength and amplitude, respectively, is about $5\,\%$ and $15\,\%$ or less. This manual method efficiently permits us to distinguish between the pattern types, as the height profiles exhibit different features in the cases of brain-like and volcano patterns. Indeed, the height profile extracted from images of samples exhibiting a brain-like pattern (Fig.~\ref{fig:profiles}A) shows typical creases with sharp and discontinuous slopes, while the height profile extracted from images of samples exhibiting a volcano pattern (Fig.~\ref{fig:profiles}B) appears generally sinusoidal, with smooth valleys and reminiscences of sharp creases on the crests.

\paragraph*{Fourier transform:~} From the AFM image of a patterned surface, we compute the 2D Fourier transform and perform a radial average of the 2D spectra. In particular, we show in Figs.~\ref{fig:profiles}C and D the averaged spectra obtained from the typical brain-like and volcano patterns showed in  Figs.~\ref{fig:profiles}A and B. The density $\rho$ is plotted as a function of the wavenumber $1/\lambda$. The wavenumber for which the density reaches a maximum gives a measurement of the dominant wavelength.

The wavelength measurement obtained with the Fourier-based method is consistent with the one obtained with the manual method previously described, however the Fourier-based method is less precise. Indeed, a typical image showing patterns with a good resolution usually shows between 5 and 20 spatial periods, which is not enough to precisely compute the Fourier transform. 

\paragraph*{Minkowski functionals:~}From the 3D AFM measurements of the surface topography we compute Minkowski functionals  \cite{Mecke1998,Fetzer2007thermalnoise}. In particular, the Minkowski area $\mathcal{A}_\mathrm{Mink}$, perimeter $\mathcal{P}_\mathrm{Mink}$, and connectivity $\chi_\mathrm{Mink}$ were measured as a function of height $z$ (and correspond to the area enclosed by an isocontour, the total  length of the isocontours, and number of connected components in an isocontour). Minkowski functionals provide a  method of distinguishing 3D patterns with different morphologies \cite{Scholz2015} since morphologically-equivalent patterns exhibit the same functional dependence on surface height.


In figures~\ref{fig:profiles}E-G, we show the Minkowski functionals computed from the AFM images showed in Fig.~\ref{fig:profiles}A and B. The Minkowski functionals clearly demonstrate that the two morphologies are geometrically distinct from one another. For our purposes, the skewness of $\chi_\mathrm{Mink}$ as a function of $z/A$, as showed in Fig.~\ref{fig:profiles}G, serves to aid classification of patterns as either brain-like or volcano-like.

\begin{figure*}
	\centering
	\includegraphics[width=0.99\linewidth]{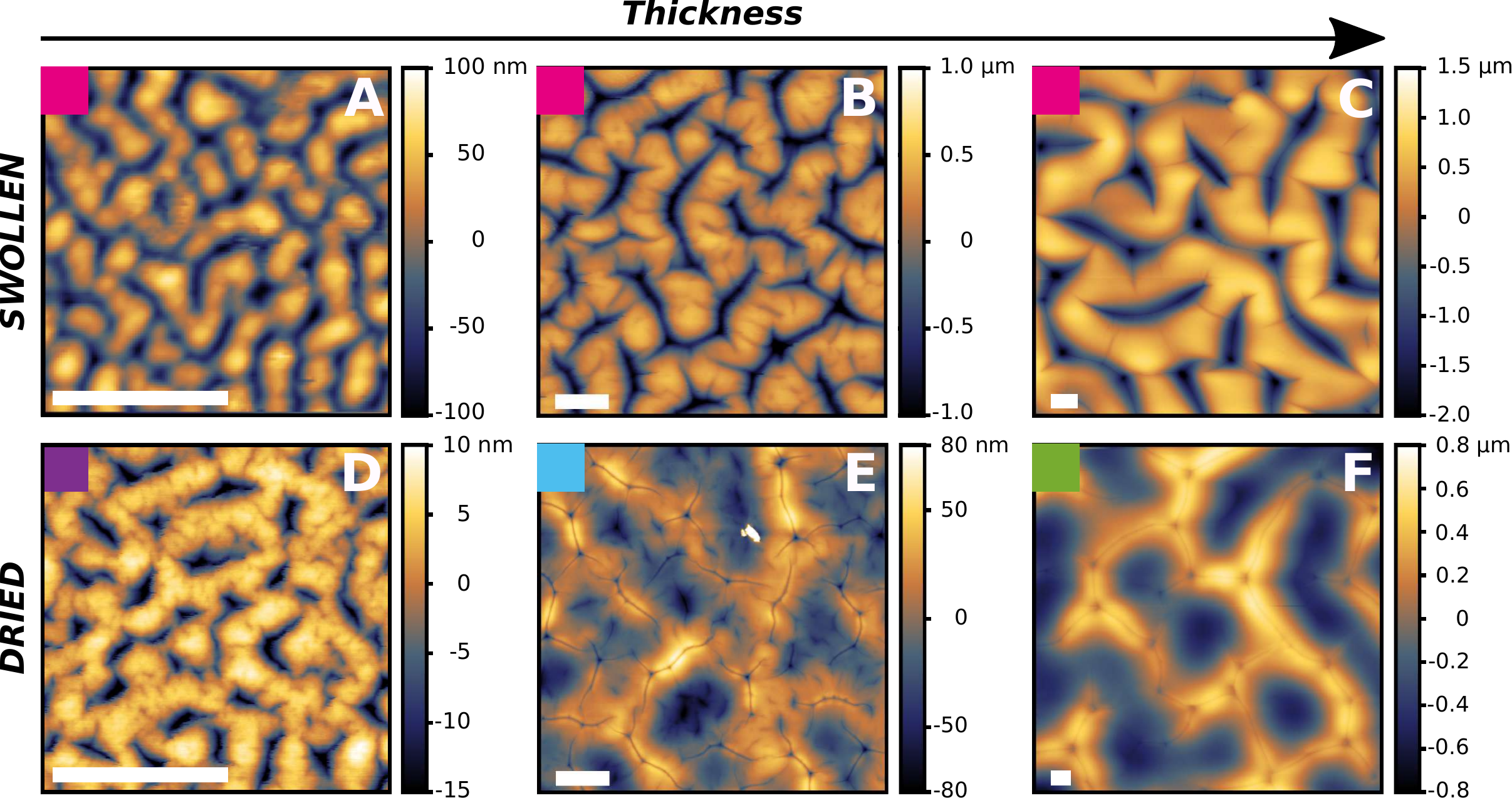}
	\caption{\small{\textbf{AFM images of dry and wet surface patterns. }\textit{ Top row: swollen state. Bottom row: dry state. Each column represents images of the same sample. The initial thickness $h$ increases from left to right. Left: $h=320\,\mathrm{nm}$. Middle: $h=950\,\mathrm{nm}$. Right: $h=4.55\,\mathrm{\mu m}$. 
	A, B, C) Films immersed in water exhibit a brain-like surface patterning. Typical swelling ratio: $S_\mathrm{R} = 3.5$.
	D) Films with initial dry thickness such that $h < l_\mathrm{ec}^\text{wet}/S_\mathrm{R} = \gamma/(G^\mathrm{wet}S_\mathrm{R})$ exhibit a brain-like surface patterning.
	E)  An example of the transition regime.
	F)  For dried films with $h>l_\mathrm{ec}^\mathrm{wet}/S_\mathrm{R}$, volcano-like patterns form. White scale bars represent 5 $\mathrm{\mu m}$ for each panel.}}}
	\label{fig:images}
\end{figure*}

In summary, the distinction between pattern types was based on the height profiles, and checked by plotting the Minkowski functionals or simply the height distribution. The measurements of the wavelength and amplitude of the instability were also made from the height profiles analysis.

\section*{Results and Discussion}
\subsection*{Results}
Typical AFM images of grafted PNIPAM films of different thicknesses are shown in Fig.~\ref{fig:images}: the surface topography and the thickness of three different samples is presented in both wet and dry conditions. During the fabrication process, such patterned morphologies appear only after the first exposition to a solvent rather than during UV irradiance (Fig.~\ref{fig:synthesis} and see SI, Sec.~II.A). The first solvent in which swelling occurs and any subsequent drying process selects once and for all the pattern observed after drying. For example, swelling in a second solvent, and repeated swelling and drying cycles, do not change the pattern (see SI. Sec.II.B). These observations suggest that surface patterning occurs in our samples as a substrate-attached hydrogel swells due to solvent imbibition, rather than because of a depth-wise gradient in crosslinking density during UV curing~ \cite{Yoon2010,Yang2010,Chandra2011,Um2021}. While we focus on the dependence of pattern morphology on film thickness, the influence of UV-irradiance time, substrate type, swelling solvent quality and vapor saturation of ambient air were also investigated as detailed in SI, Sec.~II.B. We observed that the pattern formation was independent of substrate type (silicon or glass as in SI, Sec.~I.D). The averaged rinsed-film thickness was independent of UV-irradiance time provided exposure exceeded 1.5 h (see SI, Sec.~I.D). The influence of the evaporation rate was investigated by saturating the ambient air in water vapor to slow down the evaporation. We observed that slow evaporation did not affect pattern formation (see SI, Sec.~II.B). Finally, we observed that the shape and amplitude of surface patterns are fixed after the first exposure to a rinsing solvent after UV irradiance. In particular, once formed, patterns are identical upon subsequent swelling-and-drying cycles (see SI, Sec.~II.A). These observations are consistent with descriptions of swollen gels in literature~\cite{vanLimbeek2021} and contrast with the one of compressed elastomers which exhibit reversible creases. 

Examples of the measured surface patterns shown in Fig.~\ref{fig:images} evidence a dependence of the surface topography on the initial thickness of the film. The observed surface patterns were classified by morphology as indicated by the colored squares in Fig.~\ref{fig:images}, and detailed in SI, Sec.~III. Two types of morphology with an intermediate regime are identified.
In the case of wet films shown in Figs.~\ref{fig:images}A-C and dried nanometric films in Fig.~\ref{fig:images}D, the surface patterns resemble brain-like patterns. These patterns are typical of a surface-creasing instability found in an elastomer subject to in-plane mechanical compression or a swollen hydrogel~\cite{Tanaka1987,Tanaka1992,Li1994,Li1994a,Gerardin2006,Trujillo2008,BenAmar2010,Dervaux2011,Durie2015, Chen2012,Yoon2010, Ortiz2010}. The topographies comprise smooth upper peaks separated by sharp plunging creases. The thickest film shown in Fig.~\ref{fig:images}F displays a topography that is distinct from the brain-like pattern, with broader valleys surrounding sharper peaks.  We term this morphology the ``volcano'' pattern since a small-amplitude crater is observed at the peak summits. Additional images of volcano patterns are shown in SI, Sec.~II.C.

The average thickness of both wet and dried films was measured. Surface patterns were only observed in dried and wet films with an initial thickness $h$ greater than $70\,$nm (see SI, Sec.~II.A). We note that swelling is strongly affected by the surface attachment for ultrathin PNIPAM films \cite{Li2015}. For films investigated here with $h\gtrsim 150$\,nm and 3 h irradiance time, a swelling ratio $S_\mathrm{R}$ (\textit{i.e.}~the ratio between the swollen thickness of the hydrated gel and the dry thickness) of about $3.5$ was observed. This value of swelling ratio is consistent with results from Li \textit{et al.}~\cite{Li2015} and corresponds to a mechanical strain $\epsilon= \nu(S_\mathrm{R} -1)$ of about 0.62, taking a Poisson ratio of $\nu = 0.25$ for PNIPAM~\cite{Hirotsu1991, Boon2017}. Yet, the swelling ratio decreases monotonically with thickness below $150$\,nm: this suggests that osmotic stress may be insufficient to produce a mechanical strain above the threshold of $\epsilon_\mathrm{c} = 0.35$~\cite{Hohlfeld2011,hohlfeld2008thesis, Hong2009,Trujillo2008,cai2012creasing}, corresponding to a swelling ratio of 2.4, for creasing in films with a thickness of a few tens of nanometers.

\begin{figure*}[t!]
	\centering
	\includegraphics[width=0.95\linewidth]{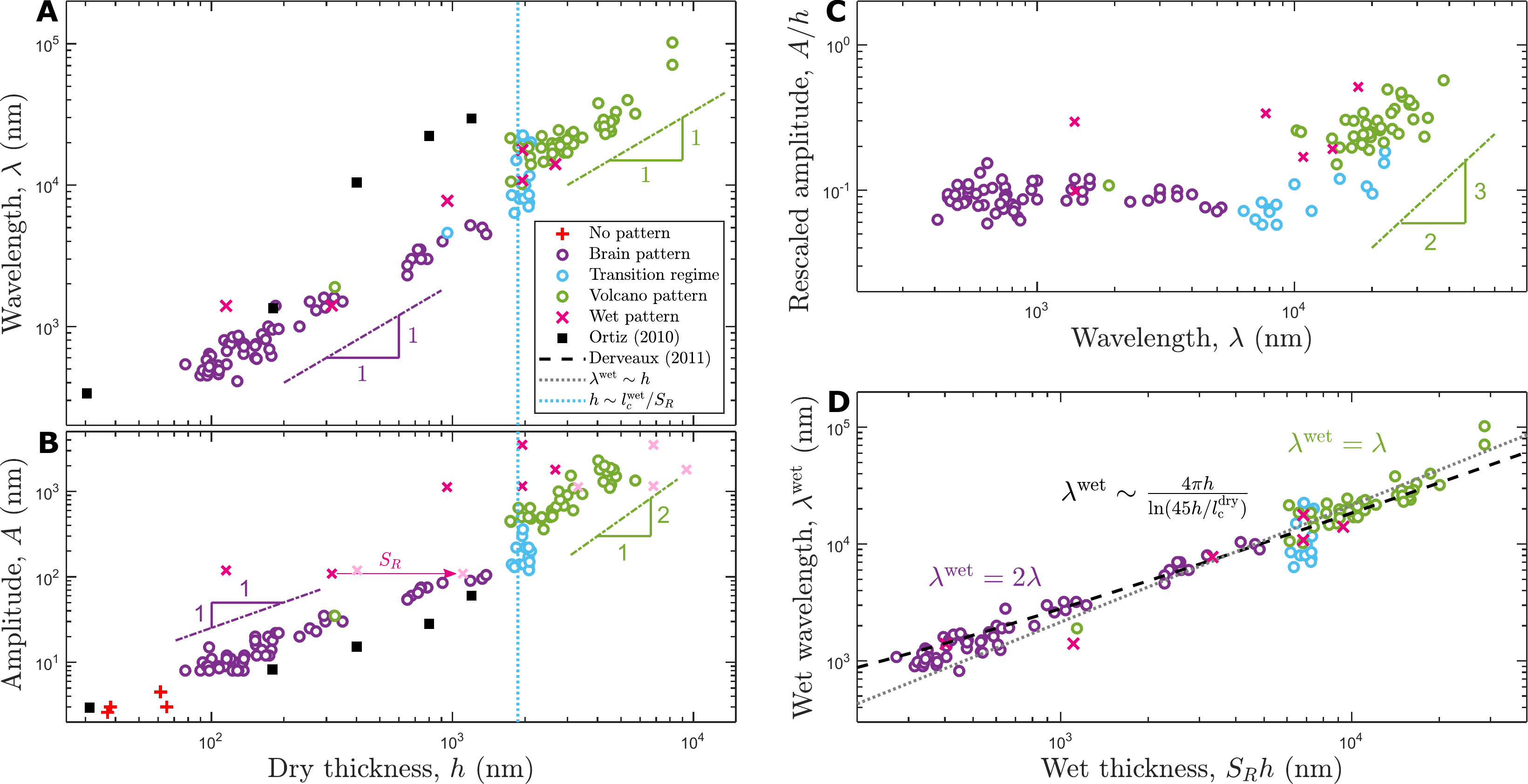}
	\caption{\small{\textbf{Wavelength and amplitude of surface patterns.} \textit{A) Wavelength as a function of dry thickness. Different morphologies correspond to different regimes (see color code). The dashed light blue line marks the transition in morphology. The slope triangles indicate power-law exponents. B) Amplitude $A$ as a function of the dry thickness $h$. Magenta crosses stand for wet amplitude, $A$, as a function of the dry thickness $h$. Light pink crosses stand for the amplitude of wet patterns as a function of swollen thickness $S_\mathrm{R}h$. The pink arrow represent the shift in thickness induced by the swelling of a factor $S_\mathrm{R}$. C) Amplitude $A$, scaled by film thickness $h$, as a function of the wavelength $\lambda$. Magenta crosses stand for wet amplitude scaled by the wet thickness $A/(S_\mathrm{R}h)$ as a function of wet wavelength $\lambda^\mathrm{wet}$. D) Expected wavelength $\lambda^\mathrm{wet}$ when swollen as a function of expected swollen thickness $S_\mathrm{R} h$. The black dashed lines represent a linear scaling (dotted line) and the scaling including a logarithmic correction} $\lambda^\mathrm{wet} \sim 4\pi h / \text{ln}(44.953 h/l_\mathrm{ec}^\mathrm{dry})$, \textit{derived by~\cite{Dervaux2011} (dashed line). In all, marker colors correspond to pattern types, as indicated in A) except for black squares which correspond to measurements of surface patterns on thin films of block-co-polymer  (PNIPAAm-co-MaBP)~\cite{Ortiz2010}. Figure adapted from~\cite{kopecz2024mechanics}, available in Open Access on \url{https://theses.hal.science/}. }}}
	\label{fig:features}
\end{figure*}

 From surface topography measured by AFM, patterns were characterized quantitatively by measuring the wavelength $\lambda$ and amplitude $A$ as well as the thickness $h$, in both wet and dry conditions. The wavelength and amplitude of the instability are respectively shown as functions of the initial, dry film thickness in Figs.~\ref{fig:features}A and B. For wet films, the wavelength increases monotonically with the initial, dry thickness. For dried films, a transition region exists between the two morphological regimes mentioned above. With increasing thickness, the wavelength is multiplied by a factor 2, at a critical thickness that we will discuss in the next section. In the thick-film regime, the wavelength of the wet pattern is comparable to that of the dried pattern, while in the thin-film regime the wavelength of the wet pattern is roughly twice the wavelength of the dried pattern (Fig.~\ref{fig:features}A). The results of Ortiz \textit{et al.} \cite{Ortiz2010}, attained with dried, thin films of PNIPAM copolymerized with methacroyloxybenzophenone (MaBP), are also included in Fig.~\ref{fig:features}A and B, for reference and are mainly consistent with our results, as we discuss in more detail in the next section.

To characterize the threshold for the formation of volcano patterns on dried films, that marks the transition between the two observed regimes, we measured the material properties of PNIPAM hydrogels in the wet state, using AFM. The Young's modulus in the swollen state was measured to be $E^{\mathrm{wet}} = 35\pm 5\,$kPa (see SI, Sec.~III.A). This value is larger but on the same order of magnitude than values from literature ($E^{\mathrm{wet}}  = 8$\,kPa), which may be due to the use of an indentation-based technique~\cite{Hashmi2009,haq2017mechanical}. Recalling that the shear modulus is expressed as $G = E/\left[2(1+\nu)\right]$, we then calculated the elastocapillary length $l_\mathrm{ec}^\mathrm{wet} = \gamma/G^\mathrm{wet} = 3.0\pm 0.5\,\mathrm{\mu}$m, with $G^\mathrm{wet}$ the shear modulus in the wet state. This elastocapillary length is based on wet mechanical properties of PNIPAM. Values were taken from literature for: the surface tension $\gamma = 41.8$\,mN/m \cite{Zhang1998} between swollen PNIPAM and air, and the Poisson’s ratio $\nu = 0.25$ \cite{Hirotsu1991,Boon2017} of PNIPAM in the swollen state and at room temperature.

Finally, from AFM images presented in Fig.~\ref{fig:images} the pattern amplitude $A$ was also measured as a function of the initial dry thickness $h$ and is presented in Fig.~\ref{fig:features}B. First, a nanometric surface roughness is measurable as for dried, ultra-thin samples that exhibit no pattern. Then for samples thick enough to exhibit a pattern ($h \gtrsim 70$ $\mathrm{nm}$), the pattern amplitude of dried films as a function of the initial film thickness switches from a linear behavior to a power law dependency at the same critical thickness as the transition in morphology, as shown in Fig.~\ref{fig:features}B. The result of Ortiz \textit{et al.} on dried films are included for comparison~\cite{Ortiz2010}: the amplitude roughly exhibits the same linear dependency in thickness. The pattern amplitude of wet films as a function of initial film thickness shows at first sight a linear behavior (see Fig.~\ref{fig:features}B, magenta crosses), but swelling induces a prefactor in thickness (see Fig.~\ref{fig:features}B, light pink crosses). 

For wet samples, the amplitude as a function of the wet thickness $S_\mathrm{R} h$ exhibits a linear scaling, similarly to the amplitude measured on dried brain-like patterns as a function of the initial dry thickness $h$, and with a similar prefactor. Thus, the ratios between dried or wet amplitude and dried or wet thickness, respectively, appears as a relevant quantity to investigate. In Fig.~\ref{fig:features}C, the ratio between amplitude and film thickness is represented as a function of the wavelength measured on each sample. The ratio between amplitude and initial film thickness shows no dependency on the wavelength for brain-like patterns, while for volcano patterns an increase in a power law is observed, with an approximate 3/2 exponent (Fig.~\ref{fig:features}C). This latter dependency is expected to saturate as the film thickness is increased even further than here, since the pattern amplitude should not become considerably larger than the undisturbed film thickness. Otherwise, more complex surface patterns such as folds may be observed as was reported elsewhere~\cite{Kim2011, Li2012, pocivavsek2008stress, oshri2017pattern, diamant2011compression}

From these observations, we conclude that surface creasing results from swelling and the wet-state properties select a particular wavelength of the observed instability. In the case of wet gels, the amplitude of the instability scales with the wet thickness. However, drying may: (i) affect the wavelength, dividing it by a factor two in the case of thin gels, (ii) modify strongly the morphology, and thus the amplitude dependency in the case of thick gels. In the next part, we discuss the thresholds in thickness for the occurrence of patterns observed on dry films, and we speculate on the origins of the morphologies. Furthermore, we compare quantitatively our experimental results on wet films with a theory proposed by Dervaux and Ben Amar \cite{Dervaux2011}. Finally, we identify key ingredients and propose a mechanism explaining the emergence of the volcano pattern observed on dry films.
\\

\subsection*{Discussion}
\emph{Thickness thresholds for morphology change:} A threshold thickness of about $70\,\mathrm{nm}$ for the creasing instability is observed and was also reported in \cite{Hayward2005}, for a given degree of cross-linking. This threshold is related to the thickness-dependent swelling ratio and thus to the volume strain. For these small thicknesses, this swelling-induced strain is not large enough to overcome the subcritical bifurcation inherent to the creasing instability \cite{Hohlfeld2011}. Then, at larger thicknesses, a second threshold thickness of about $h_\mathrm{c} \approx 2$ $\mathrm{\mu m}$ for the formation of the volcano pattern is observed on dried films. 

To rationalize this second threshold, we estimate the elastocapillary length based on the mechanical properties of the fully saturated gel, and when the swollen gel dries in contact with ambient air. This length is thus computed to be: $l_\mathrm{ec}^{\mathrm{wet}} = 3 $ $\mathrm{\mu m}$. The swelling ratio $S_\mathrm{R}$ was estimated to be about $3.5$ by AFM. Scaled by the swelling ratio, the wet elastocapillary length becomes finally $l_\mathrm{ec}^\mathrm{wet}/S_\mathrm{R} =  0.9\,\mathrm{\mu}$m.
After drying, the critical initial dry thickness at which the pattern morphology transition, from brain-like to volcanoes, is observed to be $h_\mathrm{c} \approx 2\,\mu$m. Taking into account a small overestimation of the Young's modulus due to the indentation measurement technique~\cite{haq2017mechanical,Hashmi2009}, thus an underestimation of the elastocapillary length, the dry thickness at which the transition occurs is then comparable to the wet elastocapillary lengthscale scaled by the swelling ratio. This suggests that capillary effects indeed play a role in the observed transition. Finally, the results of Ortiz \textit{et al.}~\cite{Ortiz2010} also mention a doubling of the wavelength dependency albeit at a different, yet comparable, critical thickness with our estimate of $l_\mathrm{ec}^\mathrm{wet}$. 
 
One interpretation of the correspondence between the critical thickness for morphology change and the wet elastocapillary length is as follows. The balance between surface tension and bulk elasticity determines the threshold of the surface (de)stabilization: for wet thicknesses smaller than the wet elastocapillary length, \textit{i.e.} $h^\mathrm{wet} \lesssim l_\mathrm{ec}^\mathrm{wet}$, surface tension dominates and the surface shape is not significantly affected upon drying, resulting in a creased pattern with a different spacing. For wet thicknesses larger than the wet elastocapillary length, \textit{i.e.} $h^\mathrm{wet} \gtrsim l_\mathrm{ec}^\mathrm{wet}$, bulk elasticity dominates and the surface shape is destabilized upon drying, leading to the formation of the volcano pattern.
 
The volcano morphology observed for micrometric, dried films (Fig.~\ref{fig:images}F) is distinct from that previously reported in wet millimetric hydrogels films \cite{Tanaka1987,Tanaka1992,Li1994,Li1994a,Gerardin2006}. Indeed, the patterns previously reported in thick wet hydrogel films are more akin to the brain-like patterns observed in nanometric dried films (Fig.~\ref{fig:images}D). AFM images showing the particular surface shape of the volcano pattern (Fig.~\ref{fig:images}F) are reminiscent of images shown in the work of Ortiz \textit{et.al.}~\cite{Ortiz2010}, however morphology is not discussed therein. The apparent multiscale patterning observed here is reminiscent of period doubling of wrinkle patterns which typically occur in bilayered systems comprising a thin, pre-stretched stiff film upon a soft elastomer \cite{Bowden1999, Yoo2003, Harrison2004, Stafford2006, Brau2011, Huntingdon2013, Bayley2014,Wang2015, chen2014bilayer}. We are unaware of any description of period doubling or period halving for creases in the literature. 
\\

\noindent\emph{Wavelength selection:} We have discussed the final observed morphologies and their thickness formation threshold. Starting again from the beginning of the process, we now focus on the first swelling phase. We discuss in particular the quantitative measurements of the wavelength as a function of film thickness.

Previous observations of creasing patterns in swollen hydrogel films, similar to our observations of brain-like patterns presented in Figs.~\ref{fig:images}A, B and C, have been rationalized under the theoretical framework of Dervaux and Ben Amar \cite{Dervaux2011}. The wavelength $\lambda^\mathrm{wet}$ of the brain-like patterns observed in these swollen, wet systems was found to scale with the initial dry thickness $h$ of the hydrogel: $\lambda^\mathrm{wet} \sim 4\pi h / \text{ln}(a h/l_\mathrm{ec}^\mathrm{dry})$, with $l_\mathrm{ec}^\mathrm{dry}$ the elastocapillary length computed in the initial dry state and with a numerical prefactor in the natural logarithm $a \approx 45$. This scaling implies that regularization at threshold is done by capillarity in an elastic boundary layer of initial size comparable to the dry elastocapillary length. Experimentally, crease spacings were reported to scale linearly with the thickness~\cite{Hayward2005,Gerardin2006,Ortiz2010,Durie2015}. Indeed, if the threshold for regularization is small compared to the typical sample thickness, a slight deviation from a linear scaling appears in the thinnest samples still exhibiting a pattern. 

To compare our data with the model of Dervaux and Ben Amar~\cite{Dervaux2011}, we first introduce for dried films the wavelength $\lambda^\mathrm{wet}$ that is expected in the swollen state, and that was measured for a few samples. For volcano patterns, the measured wavelength in both dried and wet conditions shows no change for a subset of samples, thus we set $\lambda^\mathrm{wet} = \lambda$. For dried brain patterns, however, we measure twice the wavelength of the dried state in the swollen state (see magenta crosses in Fig.~\ref{fig:features}A), thus we set $\lambda^\mathrm{wet} = 2\lambda$, where the factor of 2 was verified to be the best fitting value, within $2\,\%$ error. The wet wavelength $\lambda^\mathrm{wet}$ is represented as a function of the swollen thickness $S_\mathrm{R}h$ in Fig.~\ref{fig:features}D, and shows a small deviation from a linear behavior. 

The slight sublinear behavior is captured by the model of  Dervaux and Ben Amar: a fit of the data with the scaling including a logarithmic correction results in an estimation of the dry elastocapillary length of $l_\mathrm{ec}^\mathrm{dry} \approx 150$ $\mathrm{nm}$ and a corresponding shear modulus of $G^\mathrm{dry} \approx 280$ $\mathrm{kPa}$ (see SI, Sec.~III.B for the estimate precision). This estimation of shear modulus, which lies between the glassy and fully swollen ones measured here (SI, Sec.~III.A), is a value that the modulus could take after enough solvent is absorbed by the gel to cross the glass transition. In summary, with our expectation of wavelength $\lambda^\mathrm{wet}$ in the swollen state, our data for dried samples show an excellent agreement with the theoretical prediction of Dervaux and Ben Amar~\cite{Dervaux2011}. We propose the following interpretation: the geometrical expansion due to swelling is large enough to overcome the surface tension, which results in the destabilization of the free surface~\cite{Liu2019}. 

In the two previous paragraphs, we rationalized the wavelength measurements of different pattern types in the dry state by linking them with patterns observed in the wet state. Based on these observations, we conclude that the wavelength is set in the wet state, as a result of swelling. In the next paragraphs we examine how the evolution of the pattern shape during drying may proceed.
\\

\noindent\emph{Morphological evolution upon drying:} As liquid evaporates out of the gel, the film passes through a glass transition triggered by dehydration, when the polymer concentration is large enough~\cite{Leibler1993, bar2002shrinkage}. Since evaporation takes place from the free surface, at a critical solvent fraction a glassy skin layer is formed \cite{Sekimoto1993,Gennes2002,Hennessy2017, talini2023formation}. This skin layer first increases in thickness rapidly, and then expands with a diffusive dynamics~\cite{talini2023formation}. In our situation, the initial condition for drying of the swollen hydrogel films is not a flat free surface, as is the common case described in \cite{Okuzono2006}. Instead, the formation of the glassy skin layer due to evaporation takes place from an initially creased surface. We now focus on how such creasing could affect the spatial distribution of polymer within the film at the onset of evaporation. We separate the two different thickness regimes in terms of the elastocapillary length estimated above.

In the first case $h \lesssim l_\mathrm{ec}^\mathrm{wet}/S_\mathrm{R}$, surface tension effects dominate elastic stresses in the soft and swollen gel, smoothing out large-curvature features in the gel. In this regime, a drying front propagates in time, yet the film is thin enough to dry uniformly. In this situation, smaller-wavelength creases become more geometrically favorable as the scaling argument predicts $\lambda\sim h$~\cite{Dervaux2011}. However, a wavelength was already selected during swelling (Fig.~\ref{fig:3D}, first column). Following drying, we observe a new wavelength emerging with half the period. The fact that the emergent wavelength is an integer fraction of the pre-existing crease in the swollen state suggests that periodicity is preserved during drying, \textit{i.e.} the shape of the surface instability is constrained by the spatial distribution of the pre-existing creases. It seems reasonable that either the glass transition, or a crossover into the non-creasing regime of solvent fraction, is reached before a further wavelength reduction could be affected.

In a second case for films with $h \gtrsim l_\mathrm{ec}^\mathrm{wet}/S_\mathrm{R}$, we propose the following mechanism. Bulk elasticity coming from the swollen gel underneath the surface dominates over surface tension. As the gel dries, a glassy skin layer forms initially at the surface and then expands in time~\cite{talini2023formation}. The skin layer is stiffer than the still-hydrated gel underneath, by up to four orders of magnitude (see SI, Sec.~III.A for the Young's modulus measurements). Additionally, as drying generates a loss of solvent, the stiff crust would shrink, generating in-plane compressive stresses. Such stresses would pull the surface of the still-hydrated swollen gel underneath. In such a situation, a stretched, stiff and thin layer exists on a thick and soft material. Numerous works have showed that a wrinkling instability can thus appear once a critical compressive strain is exceeded~\cite{chen2014bilayer, Li2012, Kodio2017, Tallinen2013, Stafford2004, weiss2013creases,cerda2003geometry,bowden1998spontaneous,davis2011mechanics,niven2020emergence,genzer2006soft, gurmessa2017localization, pocivavsek2008stress}. The surface of the bilayer becomes destabilized, taking a sinusoidal shape, whose wavelength scales as $\lambda \sim (t_\mathrm{s}H)^{1/2}(E_\mathrm{s}/E)^{1/6}$, with $t_\mathrm{s}$ and $H$ the thicknesses, $E_\mathrm{s}$ and $E$ the Young's moduli of the skin and the substrate respectively. In our case, a wrinkling instability could thus appear on an initially-creased surface, with the spatial distribution of solvent and wavelength of wrinkles pre-selected by the creases in the swollen film. 

We estimate the thickness of the skin layer at the moment when the critical strain is exceeded using the  wavelength of the crease pattern, $t_\mathrm{s} \sim (\lambda^2/H)(E/E_\mathrm{s})^{1/3}$. We find it to be approximately $7\,\%$ of the swollen thickness of the film, using $E = E^{\mathrm{wet}} = 35\,\mathrm{kPa}$ and $E_\mathrm{s} = E^{\mathrm{dry}} = 700\,\mathrm{MPa}$ (see SI, Sec.~III.A), which is expected to be reached almost instantaneously once the formation of the skin layer is initiated~\cite{talini2023formation}. By opposition to the depth of creases and because of the extreme curvature in the region of the creases~\cite{Karpitschka2017}, the bumped regions are less constrained and thus shrink more during evaporation, which may also occur preferentially from the peaks of the bumps than in the vicinity of the creases~\cite{moore2021nascent}, possibly creating an inversion of the topography. Indeed we observe the resulting surface morphology to be characterized by a globally-sinusoidal profile, with scars of creases present at the crests. Thus the wavelength of the dried volcano pattern is prescribed by the swollen creased patterns initiated in the wet state. 

In summary, swelling and drying hydrogels may experience elastic instabilities that affect the surface topography. During swelling the originally flat surface is deformed by an in-plane compressive stress, resulting in the formation of creases. During drying, depending on the balance between surface tension and bulk elasticity, the formation of a thin, stiff and glassy skin layer on the still-hydrated and soft gel may significantly affect the final surface morphology of the dried film. The final surface morphology of dried films can therefore be a brain-like or a volcano pattern and is determined by this drying process, depending on the thickness of the wet film before drying commences. \\

\section*{Conclusions}
We have investigated the irreversible deformation of grafted PNIPAM films, of nanometric and micrometric thickness, caused by imbibition-induced swelling and evaporation-driven shrinking. Relatively thin, nanometric films undergo a swelling instability that leads to a brain-like pattern, typical of surface creasing instabilities and with a set spacing. When drying, the surface morphology keeps the creased shape although the spatial frequency of creases is doubled as compared to the wet state. The new pattern is then locked-in as the film shrinks (through the glass transition) into a dried film. The typical amplitude of these surface patterns with sub-micrometric wavelengths ($\lambda \leq 1\,\mu$m) is consistently $\sim 10\%$ of the film thickness, independent of the wavelength (see Fig.~\ref{fig:features}C). 
Thicker, micrometric films also develop surface creases because of in-plane compressive forces that result from constrained swelling due to surface attachment. However, drying is known to induce the formation of a glassy skin layer. We propose that this glassy layer is responsible for a subsequent change in topographic morphology, rendering it distinct from the brain-like pattern observed in the wet state. \\

\noindent\textbf{Funding:} %
The authors acknowledge financial support from the CNRS, and from the Agence Nationale de la Recherche (ANR) under the CoPinS (ANR-19-CE06-0021) grant, the Institut Pierre-Gilles de Gennes (Equipex ANR-10-EQPX-34 and Labex ANR-10-LABX- 31) and PSL Research University (Idex ANR-10-IDEX-0001-02). They also acknowledge support from the European Union through the European Research Council under EMetBrown (ERC-CoG-101039103) grant. They finally aknowledge the Royal Society (URF/R1/211730).

\noindent\textbf{Conflicts of Interest:} The authors declare no conflict of interest and the funders had no role in the design of the study; in the collection, analyses, or interpretation of data; in the writing of the manuscript, or in the decision to publish the results.

\noindent\textbf{Supporting Information:} Additional experimental details, materials, methods, and peripheral observations.

\noindent\textbf{Data Availability:} The reported experimental data are available in the SI.

\begin{acknowledgments}
The authors are more than grateful to Marjan Abdorahim for sharing her strong expertise on hydrogels.
The authors gratefully acknowledge Patrick Tabeling, Ousmane Kodio, Ken Sekimoto, Grae Worster, Joseph Webber, Noushine Shahidzadeh and Romane Le Dizès Castell for insightful comments and fruitful discussions. The AFM work benefited from EUSMI TA support and expertise of Dr. Peisker from Nanosurf. The experimental work benefited from the technical contribution of the joint service unit CNRS UAR 3750. Finally, the authors thank the Soft Matter Collaborative Research Unit, Frontier Research Center for Advanced Material and Life Science, Faculty of Advanced Life Science at Hokkaido University, Sapporo, Japan.
\end{acknowledgments}

\section*{References}

\begin{thebibliography}{0}%
\makeatletter
\providecommand \@ifxundefined [1]{%
 \@ifx{#1\undefined}
}%
\providecommand \@ifnum [1]{%
 \ifnum #1\expandafter \@firstoftwo
 \else \expandafter \@secondoftwo
 \fi
}%
\providecommand \@ifx [1]{%
 \ifx #1\expandafter \@firstoftwo
 \else \expandafter \@secondoftwo
 \fi
}%
\providecommand \natexlab [1]{#1}%
\providecommand \enquote  [1]{``#1''}%
\providecommand \bibnamefont  [1]{#1}%
\providecommand \bibfnamefont [1]{#1}%
\providecommand \citenamefont [1]{#1}%
\providecommand \href@noop [0]{\@secondoftwo}%
\providecommand \href [0]{\begingroup \@sanitize@url \@href}%
\providecommand \@href[1]{\@@startlink{#1}\@@href}%
\providecommand \@@href[1]{\endgroup#1\@@endlink}%
\providecommand \@sanitize@url [0]{\catcode `\\12\catcode `\$12\catcode
  `\&12\catcode `\#12\catcode `\^12\catcode `\_12\catcode `\%12\relax}%
\providecommand \@@startlink[1]{}%
\providecommand \@@endlink[0]{}%
\providecommand \url  [0]{\begingroup\@sanitize@url \@url }%
\providecommand \@url [1]{\endgroup\@href {#1}{\urlprefix }}%
\providecommand \urlprefix  [0]{URL }%
\providecommand \Eprint [0]{\href }%
\providecommand \doibase [0]{https://doi.org/}%
\providecommand \selectlanguage [0]{\@gobble}%
\providecommand \bibinfo  [0]{\@secondoftwo}%
\providecommand \bibfield  [0]{\@secondoftwo}%
\providecommand \translation [1]{[#1]}%
\providecommand \BibitemOpen [0]{}%
\providecommand \bibitemStop [0]{}%
\providecommand \bibitemNoStop [0]{.\EOS\space}%
\providecommand \EOS [0]{\spacefactor3000\relax}%
\providecommand \BibitemShut  [1]{\csname bibitem#1\endcsname}%
\let\auto@bib@innerbib\@empty
\end{thebibliography}%


\begin{thebibliography}{100}

\bibitem{Russell2002}
T.~P. Russell, ``Surface-responsive materials,'' {\em Science}, vol.~297,
  no.~5583, pp.~964--967, 2002.

\bibitem{ahn2008stimuli}
S.-k. Ahn, R.~M. Kasi, S.-C. Kim, N.~Sharma, and Y.~Zhou, ``Stimuli-responsive
  polymer gels,'' {\em Soft Matter}, vol.~4, no.~6, pp.~1151--1157, 2008.

\bibitem{Li1994a}
C.~Li, Z.~Hu, and Y.~Li, ``Temperature and time dependencies of surface
  patterns in constrained ionic n‐isopropylacrylamide gels,'' {\em J. Chem.
  Phys.}, vol.~100, no.~6, pp.~4645--4652, 1994.

\bibitem{kim2010dynamic}
J.~Kim, J.~Yoon, and R.~C. Hayward, ``Dynamic display of biomolecular patterns
  through an elastic creasing instability of stimuli-responsive hydrogels,''
  {\em Nat. Mater.}, vol.~9, no.~2, pp.~159--164, 2010.

\bibitem{Yoon2010}
J.~Yoon, J.~Kim, and R.~C. Hayward, ``Nucleation{,} growth{,} and hysteresis of
  surface creases on swelled polymer gels,'' {\em Soft Matter}, vol.~6,
  pp.~5807--5816, 2010.

\bibitem{Beebe2000}
D.~Beebe, J.~Moore, J.~Bauer, Q.~Yu, R.~H. Liu, C.~Devadoss, and B.-H. Jo,
  ``Free-evolution kinetics in a high-swelling polymeric hydrogel,'' {\em
  Nature}, vol.~404, pp.~588--590, 2000.

\bibitem{Idota2005}
N.~Idota, A.~Kikuchi, J.~Kobayashi, K.~Sakai, and T.~Okano, ``Microfluidic
  valves comprising nanolayered thermoresponsive polymer-grafted capillaries,''
  {\em Adv. Mater.}, vol.~17, no.~22, pp.~2723--2727.

\bibitem{DEramo2018}
L.~{D'E}ramo, B.~Chollet, M.~Leman, E.~Martwong, M.~Li, H.~Geisler, J.~Dupire,
  M.~Kerdraon, C.~Vergne, F.~Monti, Y.~Tran, and P.~Tabeling, ``Microfluidic
  actuators based on temperature-responsive hydrogels,'' {\em Microsys.
  Nanoeng.}, vol.~4, no.~1, p.~17069, 2018.

\bibitem{Li2015}
M.~Li, B.~Bresson, F.~Cousin, C.~Fr\'{e}tigny, and Y.~Tran, ``Submicrometric
  films of surface-attached polymer network with temperature-responsive
  properties,'' {\em Langmuir}, vol.~31, p.~11516–11524, 2015.

\bibitem{winnik1990fluorescence}
F.~M. Winnik, ``Fluorescence studies of aqueous solutions of poly
  (n-isopropylacrylamide) below and above their lcst,'' {\em Macromolecules},
  vol.~23, no.~1, pp.~233--242, 1990.

\bibitem{heskins1968solution}
M.~Heskins and J.~E. Guillet, ``Solution properties of poly
  (n-isopropylacrylamide),'' {\em J. Macromol. Sci. Chem.}, vol.~2, no.~8,
  pp.~1441--1455, 1968.

\bibitem{haq2017mechanical}
M.~A. Haq, Y.~Su, and D.~Wang, ``Mechanical properties of pnipam based
  hydrogels: A review,'' {\em Mater. Sci. Eng. C}, vol.~70, pp.~842--855, 2017.

\bibitem{Peppas2006}
N.~Peppas, J.~Hilt, A.~Khademhosseini, and R.~Langer, ``Hydrogels in biology
  and medicine: From molecular principles to bionanotechnology,'' {\em Adv.
  Mater.}, vol.~18, no.~11, pp.~1345--1360, 2006.

\bibitem{Niloofar2016}
N.~E., M.~Abdorahim, and A.~Simchi, ``Smart polymeric hydrogels for cartilage
  tissue engineering: A review on the chemistry and biological functions,''
  {\em Biomacromolecules}, vol.~17, 10 2016.

\bibitem{Trujillo2008}
V.~Trujillo, J.~Kim, and R.~C. Hayward, ``Creasing instability of
  surface-attached hydrogels,'' {\em Soft Matter}, vol.~4, pp.~564--569, 2008.

\bibitem{Hong2009}
W.~Hong, X.~Zhao, and Z.~Suo, ``Formation of creases on the surfaces of
  elastomers and gels,'' {\em Appl. Phys. Lett.}, vol.~95, no.~11, p.~111901,
  2009.

\bibitem{Ortiz2010}
O.~Ortiz, A.~Vidyasagar, J.~Wang, and R.~Toomey, ``Surface instabilities in
  ultrathin, cross-linked poly(n-isopropylacrylamide) coatings,'' {\em
  Langmuir}, vol.~26, p.~17489–17494, 2010.

\bibitem{Chen2014}
D.~Chen, J.~Yoon, D.~Chandra, A.~J. Crosby, and R.~C. Hayward,
  ``Stimuli-responsive buckling mechanics of polymer films,'' {\em J. Polym.
  Sci., Part B: Polym. Phys.}, vol.~52, no.~22, pp.~1441--1461, 2014.

\bibitem{Ju2022}
J.~Ju, K.~Sekimoto, L.~Cipelletti, C.~Creton, and T.~Narita, ``Heterogeneous
  nucleation of creases in swelling polymer gels,'' {\em Phys. Rev. E},
  vol.~105, p.~034504, Mar 2022.

\bibitem{Ciarletta2018}
P.~Ciarletta, ``Matched asymptotic solution for crease nucleation in soft
  solids,'' {\em Nat. Commun.}, vol.~9, p.~496, 2018.

\bibitem{cai2012creasing}
S.~Cai, D.~Chen, Z.~Suo, and R.~C. Hayward, ``Creasing instability of elastomer
  films,'' {\em Soft Matter}, vol.~8, no.~5, pp.~1301--1304, 2012.

\bibitem{Hohlfeld2011}
E.~Hohlfeld and L.~Mahadevan, ``Unfolding the sulcus,'' {\em Phys. Rev. Lett.},
  vol.~106, p.~105702, 2011.

\bibitem{Chen2012}
D.~Chen, S.~Cai, Z.~Suo, and R.~C. Hayward, ``Surface energy as a barrier to
  creasing of elastomer films: An elastic analogy to classical nucleation,''
  {\em Phys. Rev. Lett.}, vol.~109, p.~038001, 2012.

\bibitem{Suematsu1990}
N.~Suematsu, K.~Sekimoto, and K.~Kawasaki, ``Three-dimensional computer
  modeling for pattern formation on the surface of an expanding polymer gel,''
  {\em Phys. Rev. A}, vol.~41, pp.~5751--5754, 1990.

\bibitem{Li2012}
B.~Li, Y.-P. Cao, X.-Q. Feng, and H.~Gao, ``Mechanics of morphological
  instabilities and surface wrinkling in soft materials: a review,'' {\em Soft
  Matter}, vol.~8, pp.~5728--5745, 2012.

\bibitem{Tanaka1987}
T.~Tanaka, S.-T. Sun, Y.~Hirokawa, S.~Katayama, J.~Kucera, Y.~Hirose, and
  T.~Amiya, ``Mechanical instability of gels at the phase transition,'' {\em
  Nature}, vol.~325, pp.~796--798, 1987.

\bibitem{Tanaka1992}
H.~Tanaka, H.~Tomita, A.~Takasu, T.~Hayashi, and T.~Nishi, ``Morphological and
  kinetic evolution of surface patterns in gels during the swelling process:
  {E}vidence of dynamic pattern ordering,'' {\em Phys. Rev. Lett.}, vol.~68,
  pp.~2794--2797, 1992.

\bibitem{Li1994}
Y.~Li, C.~Li, and Z.~Hu, ``Pattern formation of constrained acrylamide/sodium
  acrylate copolymer gels in acetone/water mixture,'' {\em J. Chem. Phys.},
  vol.~100, no.~6, pp.~4637--4644, 1994.

\bibitem{Gerardin2006}
H.~G\'{e}rardin, A.~Buguin, and F.~Brochard-Wyart, {\em Pattern formation in
  anti-fog and anti-frost polymer coatings}.
\newblock PhD thesis, 2006.

\bibitem{Durie2015}
K.~Durie, M.~J. Razavi, X.~Wang, and J.~Locklin, ``Nanoscale surface creasing
  induced by post-polymerization modification,'' {\em ACS Nano}, vol.~9,
  p.~10961–10969, 2015.

\bibitem{Sheppard1918}
S.~E. Sheppard and F.~A. Elliott, ``The reticulation of gelatine,'' {\em J.
  Ind. Eng. Chem.}, vol.~10, no.~9, pp.~727--732, 1918.

\bibitem{southern1965effect}
E.~Southern and A.~Thomas, ``Effect of constraints on the equilibrium swelling
  of rubber vulcanizates,'' {\em J. Polym. Sci., Part A: Gen. Pap.}, vol.~3,
  no.~2, pp.~641--646, 1965.

\bibitem{Biot1963}
M.~A. Biot, ``Surface instability of rubber in compression,'' {\em Appl. Sci.
  Res. A}, vol.~12, pp.~168--182, 1963.

\bibitem{mora2011surface}
S.~Mora, M.~Abkarian, H.~Tabuteau, and Y.~Pomeau, ``Surface instability of soft
  solids under strain,'' {\em Soft Matter}, vol.~7, no.~22, pp.~10612--10619,
  2011.

\bibitem{hohlfeld2008thesis}
E.~B. Hohlfeld, ``Creasing, point-bifurcations, and the spontaneous breakdown
  of scale-invariance,'' {\em Ph. D. Thesis}, 2008.

\bibitem{Cao2012}
Y.~Cao and J.~W. Hutchinson, ``From wrinkles to creases in elastomers: the
  instability and imperfection-sensitivity of wrinkling,'' {\em Proc. R. Soc.
  A.}, vol.~468, p.~94–115, 2012.

\bibitem{wong2010surface}
W.~Wong, T.~Guo, Y.~Zhang, and L.~Cheng, ``Surface instability maps for soft
  materials,'' {\em Soft Matter}, vol.~6, no.~22, pp.~5743--5750, 2010.

\bibitem{Tallinen2013}
T.~Tallinen, J.~S. Biggins, and L.~Mahadevan, ``Surface sulci in squeezed soft
  solids,'' {\em Phys. Rev. Lett.}, vol.~110, p.~024302, 2013.

\bibitem{Guvendiren2010}
M.~Guvendiren, J.~A. Burdick, and S.~Yang, ``Solvent induced transition from
  wrinkles to creases in thin film gels with depth-wise crosslinking
  gradients,'' {\em Soft Matter}, vol.~6, pp.~5795--5801, 2010.

\bibitem{drummond1988surface}
W.~R. Drummond, M.~L. Knight, M.~L. Brannon, and N.~A. Peppas, ``Surface
  instabilities during swelling of ph-sensitive hydrogels,'' {\em J. Controlled
  Release}, vol.~7, no.~2, pp.~181--183, 1988.

\bibitem{ghatak2007kink}
A.~Ghatak and A.~L. Das, ``Kink instability of a highly deformable elastic
  cylinder,'' {\em Phys. Rev. Lett.}, vol.~99, no.~7, p.~076101, 2007.

\bibitem{chen2014bilayer}
D.~Chen, L.~Jin, Z.~Suo, and R.~C. Hayward, ``Controlled formation and
  disappearance of creases,'' {\em Mater. Horiz.}, vol.~1, no.~2, pp.~207--213,
  2014.

\bibitem{deegan2000pattern}
R.~D. Deegan, ``Pattern formation in drying drops,'' {\em Phys. Rev. E},
  vol.~61, no.~1, p.~475, 2000.

\bibitem{zang2019evaporation}
D.~Zang, S.~Tarafdar, Y.~Y. Tarasevich, M.~D. Choudhury, and T.~Dutta,
  ``Evaporation of a droplet: From physics to applications,'' {\em Phys. Rep.},
  vol.~804, pp.~1--56, 2019.

\bibitem{ozawa2005modeling}
K.~Ozawa, E.~Nishitani, and M.~Doi, ``Modeling of the drying process of liquid
  droplet to form thin film,'' {\em Jpn. J. Appl. Phys.}, vol.~44, no.~6R,
  p.~4229, 2005.

\bibitem{pauchard2003stable}
L.~Pauchard and C.~Allain, ``Stable and unstable surface evolution during the
  drying of a polymer solution drop,'' {\em Phys. Rev. E}, vol.~68, no.~5,
  p.~052801, 2003.

\bibitem{de2001instabilities}
P.-G. De~Gennes, ``Instabilities during the evaporation of a film: Non-glassy
  polymer+ volatile solvent,'' {\em Eur. Phys. J. E}, vol.~6, pp.~421--424,
  2001.

\bibitem{de2002solvent}
P.-G. De~Gennes, ``Solvent evaporation of spin cast films:“crust”
  effects,'' {\em Eur. Phys. J. E}, vol.~7, pp.~31--34, 2002.

\bibitem{routh2013drying}
A.~F. Routh, ``Drying of thin colloidal films,'' {\em Rep. Prog. Phys.},
  vol.~76, no.~4, p.~046603, 2013.

\bibitem{okuzono2006simple}
T.~Okuzono, K.~Ozawa, and M.~Doi, ``Simple model of skin formation caused by
  solvent evaporation in polymer solutions,'' {\em Phys. Rev. Lett.}, vol.~97,
  no.~13, p.~136103, 2006.

\bibitem{bornside1989spin}
D.~Bornside, C.~Macosko, and L.~Scriven, ``Spin coating: One-dimensional
  model,'' {\em J. Appl. Phys.}, vol.~66, no.~11, pp.~5185--5193, 1989.

\bibitem{fu2019differential}
N.~Fu, M.~Yu, and X.~D. Chen, ``A differential shrinkage approach for
  evaluating particle formation behavior during drying of sucrose, lactose,
  mannitol, skim milk, and other solid-containing droplets,'' {\em Dry.
  Technol.}, vol.~37, no.~8, pp.~941--949, 2019.

\bibitem{Bertrand2016}
T.~Bertrand, J.~Peixinho, S.~Mukhopadhyay, and C.~W. MacMinn, ``Dynamics of
  swelling and drying in a spherical gel,'' {\em Phys. Rev. Applied}, vol.~6,
  p.~064010, 2016.

\bibitem{Etzold2021}
M.~A. Etzold, P.~F. Linden, and M.~G. Worster, ``Transpiration through
  hydrogels,'' {\em J. Fluid Mech}, vol.~925, no.~A8, pp.~1000--1003, 2021.

\bibitem{Engelsberg2013}
M.~Engelsberg and W.~Barros, ``Free-evolution kinetics in a high-swelling
  polymeric hydrogel,'' {\em Phys. Rev. E}, vol.~88, p.~062602, 2013.

\bibitem{deegan2000contact}
R.~D. Deegan, O.~Bakajin, T.~F. Dupont, G.~Huber, S.~R. Nagel, and T.~A.
  Witten, ``Contact line deposits in an evaporating drop,'' {\em Phys. Rev. E},
  vol.~62, no.~1, p.~756, 2000.

\bibitem{hennessy2017minimal}
M.~G. Hennessy, G.~L. Ferretti, J.~T. Cabral, and O.~K. Matar, ``A minimal
  model for solvent evaporation and absorption in thin films,'' {\em J.
  Colloid. Interface Sci.}, vol.~488, pp.~61--71, 2017.

\bibitem{deegan1997capillary}
R.~D. Deegan, O.~Bakajin, T.~F. Dupont, G.~Huber, S.~R. Nagel, and T.~A.
  Witten, ``Capillary flow as the cause of ring stains from dried liquid
  drops,'' {\em Nature}, vol.~389, no.~6653, pp.~827--829, 1997.

\bibitem{pauchard2003buckling}
L.~Pauchard and C.~Allain, ``Buckling instability induced by polymer solution
  drying,'' {\em Europhys. Lett.}, vol.~62, no.~6, p.~897, 2003.

\bibitem{larson2014transport}
R.~G. Larson, ``Transport and deposition patterns in drying sessile droplets,''
  {\em AIChE J.}, vol.~60, no.~5, pp.~1538--1571, 2014.

\bibitem{hu2005analysis}
H.~Hu and R.~G. Larson, ``Analysis of the microfluid flow in an evaporating
  sessile droplet,'' {\em Langmuir}, vol.~21, no.~9, pp.~3963--3971, 2005.

\bibitem{hu2005analysisMarangoni}
H.~Hu and R.~G. Larson, ``Analysis of the effects of marangoni stresses on the
  microflow in an evaporating sessile droplet,'' {\em Langmuir}, vol.~21,
  no.~9, pp.~3972--3980, 2005.

\bibitem{poulard2007control}
C.~Poulard and P.~Damman, ``Control of spreading and drying of a polymer
  solution from marangoni flows,'' {\em Europhys. Lett.}, vol.~80, no.~6,
  p.~64001, 2007.

\bibitem{pauchard2003mechanical}
L.~Pauchard and C.~Allain, ``Mechanical instability induced by complex liquid
  desiccation,'' {\em C. R. Phys.}, vol.~4, no.~2, pp.~231--239, 2003.

\bibitem{zhou2017structure}
J.~Zhou, X.~Man, Y.~Jiang, and M.~Doi, ``Structure formation in soft-matter
  solutions induced by solvent evaporation,'' {\em Adv. Mater.}, vol.~29,
  no.~45, p.~1703769, 2017.

\bibitem{Vermant_2005}
J.~Vermant and M.~J. Solomon, ``Flow-induced structure in colloidal
  suspensions,'' {\em J. Condens. Matter Phys.}, vol.~17, p.~R187, jan 2005.

\bibitem{chen2009arched}
L.~Chen and J.~R. Evans, ``Arched structures created by colloidal droplets as
  they dry,'' {\em Langmuir}, vol.~25, no.~19, pp.~11299--11301, 2009.

\bibitem{mcgraw2010plateau}
J.~D. McGraw, J.~Li, D.~L. Tran, A.-C. Shi, and K.~Dalnoki-Veress,
  ``Plateau-rayleigh instability in a torus: formation and breakup of a polymer
  ring,'' {\em Soft Matter}, vol.~6, no.~6, pp.~1258--1262, 2010.

\bibitem{mcgraw2011dynamics}
J.~D. McGraw, I.~D. Rowe, M.~Matsen, and K.~Dalnoki-Veress, ``Dynamics of
  interacting edge defects in copolymer lamellae,'' {\em Eur. Phys. J. E},
  vol.~34, pp.~1--7, 2011.

\bibitem{xie2021delamination}
K.~Xie, A.~Glasser, S.~Shinde, Z.~Zhang, J.-M. Rampnoux, A.~Maali, E.~Cloutet,
  G.~Hadziioannou, and H.~Kellay, ``Delamination and wrinkling of flexible
  conductive polymer thin films,'' {\em Adv. Func. Mater.}, vol.~31, no.~21,
  p.~2009039, 2021.

\bibitem{Saintyves2023}
B.~Saintyves, R.~Pic, L.~Mahadevan, and I.~Bischofberger, ``Evaporation-driven
  cellular patterns in confined hyperelastic hydrogels,'' {\em Phys. Rev.
  Lett.}, vol.~131, p.~118202, Sep 2023.

\bibitem{kolb2001click}
H.~C. Kolb, M.~Finn, and K.~B. Sharpless, ``Click chemistry: diverse chemical
  function from a few good reactions,'' {\em Angew. Chem. Int. Ed.}, vol.~40,
  no.~11, pp.~2004--2021, 2001.

\bibitem{chollet2016tailoring}
B.~Chollet, L.~D’eramo, E.~Martwong, M.~Li, J.~Macron, T.~Q. Mai,
  P.~Tabeling, and Y.~Tran, ``Tailoring patterns of surface-attached
  multiresponsive polymer networks,'' {\em ACS Applied Materials \&
  Interfaces}, vol.~8, no.~37, pp.~24870--24879, 2016.

\bibitem{benjamin2016multiscale}
C.~Benjamin, L.~Mengxing, M.~Ekkachai, B.~Bruno, F.~Christian, T.~Patrick, and
  T.~Yvette, ``Multiscale surface-attached hydrogel thin films with tailored
  architecture,'' 2016.

\bibitem{kopecz2024mechanics}
C.~Kopecz-Muller, {\em Mechanics of hydrogel films: swelling-induced
  instabilities, finite-size effects, and contactless rheology to
  indentation-induced dehydration}.
\newblock PhD thesis, Universit{\'e} de Bordeaux, 2024.

\bibitem{BenAmar2010}
M.~Ben~Amar and P.~Ciarletta, ``Swelling instability of surface-attached gels
  as a model of soft tissue growth under geometric constraints,'' {\em J. Mech.
  Phys. Solids}, vol.~58, pp.~935--954, 07 2010.

\bibitem{Dervaux2011}
J.~Dervaux and M.~{Ben Amar}, ``Buckling condensation in constrained growth,''
  {\em J. Mech. Phys. Solids}, vol.~59, no.~3, pp.~538--560, 2011.

\bibitem{Kang2010}
M.~K. Kang and R.~Huang, ``Swell-induced surface instability of confined
  hydrogel layers on substrates,'' {\em J. Mech. Phys. Solids}, vol.~58,
  no.~10, pp.~1582--1598, 2010.

\bibitem{Mora2011}
S.~Mora, M.~Abkarian, H.~Tabuteau, and Y.~Pomeau, ``Surface instability of soft
  solids under strain,'' {\em Soft Matter}, vol.~7, pp.~10612--10619, 2011.

\bibitem{Liu2019}
Q.~Liu, T.~Ouchi, L.~Jin, R.~Hayward, and Z.~Suo, ``Elastocapillary crease,''
  {\em Phys. Rev. Lett.}, vol.~122, p.~098003, 2019.

\bibitem{vanLimbeek2021}
M.~A.~J. {van Limbeek}, M.~H. Essink, A.~Pandey, J.~H. Snoeijer, and
  S.~Karpitschka, ``Pinning-induced folding-unfolding asymmetry in adhesive
  creases,'' {\em Phys. Rev. Lett.}, vol.~127, p.~028001, 2021.

\bibitem{Mecke1998}
K.~R. Mecke, ``Integral geometry in statistical physics,'' {\em Int. J. Modern
  Phys. B}, vol.~12, no.~09, pp.~861--899, 1998.

\bibitem{Fetzer2007thermalnoise}
R.~Fetzer, M.~Rauscher, R.~Seemann, K.~Jacobs, and K.~Mecke, ``Thermal noise
  influences fluid flow in thin films during spinodal dewetting,'' {\em Phys.
  Rev. Lett.}, vol.~99, p.~114503, Sep 2007.

\bibitem{Scholz2015}
C.~Scholz, G.~E. Schr\"{o}der-Turk, and K.~Mecke, ``Pattern-fluid
  interpretation of chemical turbulence,'' {\em Phys. Rev. E.}, vol.~91, no.~4,
  p.~042907, 2015.

\bibitem{Yang2010}
S.~Yang, K.~Khare, and P.-C. Lin, ``Harnessing surface wrinkle patterns in soft
  matter,'' {\em Adv. Funct. Mater.}, vol.~20, p.~2550–2564, 2010.

\bibitem{Chandra2011}
D.~Chandra and A.~Crosby, ``Self‐wrinkling of uv‐cured polymer films,''
  {\em Adv. Mater.}, vol.~23, pp.~3441--3445, 2011.

\bibitem{Um2021}
E.~Um, Y.-K. Cho, and J.~Jeong, ``Spontaneous wrinkle formation on hydrogel
  surfaces using photoinitiator diffusion from oil-water interface,'' {\em ACS
  Appl. Mater. Interfaces.}, vol.~13, pp.~15837--15846, 2021.

\bibitem{Hirotsu1991}
S.~Hirotsu, ``Softening of bulk modulus and negative poisson’s ratio near the
  volume phase transition of polymer gels,'' {\em J. Chem. Phys.}, vol.~94,
  no.~5, pp.~3949--3957, 1991.

\bibitem{Boon2017}
N.~Boon and P.~Schurtenberger, ``Swelling of micro-hydrogels with a crosslinker
  gradient,'' {\em Phys. Chem. Chem. Phys.}, vol.~19, pp.~23740--23746, 2017.

\bibitem{Hashmi2009}
S.~M. Hashmi and E.~R. Dufresne, ``Mechanical properties of individual microgel
  particles through the deswelling transition,'' {\em Soft Matter}, vol.~5,
  pp.~3682--3688, 2009.

\bibitem{Zhang1998}
J.~Zhang and R.~Pelton, ``The surface tension of aqueous
  poly(n-isopropylacrylamide-co-acrylamide),'' {\em J. Polymer Sci. A},
  vol.~37, pp.~2137--2143, 1998.

\bibitem{Kim2011}
P.~Kim, M.~Abkarian, and H.~Stone, ``Hierarchical folding of elastic membranes
  under biaxial compressive stress,'' {\em Nature Mater.}, vol.~10,
  p.~952–957, 2011.

\bibitem{pocivavsek2008stress}
L.~Pocivavsek, R.~Dellsy, A.~Kern, S.~Johnson, B.~Lin, K.~Y.~C. Lee, and
  E.~Cerda, ``Stress and fold localization in thin elastic membranes,'' {\em
  Science}, vol.~320, no.~5878, pp.~912--916, 2008.

\bibitem{oshri2017pattern}
O.~Oshri and H.~Diamant, ``Pattern transitions in a compressible floating
  elastic sheet,'' {\em Physical Chemistry Chemical Physics}, vol.~19, no.~35,
  pp.~23817--23824, 2017.

\bibitem{diamant2011compression}
H.~Diamant and T.~A. Witten, ``Compression induced folding of a sheet: An
  integrable system,'' {\em Physical review letters}, vol.~107, no.~16,
  p.~164302, 2011.

\bibitem{Hayward2005}
R.~C. Hayward, B.~F. Chmelka, and E.~J. Kramer, ``Template cross-linking
  effects on morphologies of swellable block copolymer and mesostructured
  silica thin films,'' {\em Macromolecules}, vol.~38, p.~7768–7783, 2005.

\bibitem{Bowden1999}
N.~Bowden, W.~T.~S. Huck, K.~E. Paul, and G.~M. Whitesides, ``The controlled
  formation of ordered, sinusoidal structures by plasma oxidation of an
  elastomeric polymer,'' {\em App. Phys. Lett.}, vol.~75, no.~17,
  pp.~2557--2559, 1999.

\bibitem{Yoo2003}
P.~J. Yoo and H.~H. Lee, ``Evolution of a stress-driven pattern in thin bilayer
  films: spinodal wrinkling,'' {\em Phys. Rev. Lett}, vol.~91, p.~154502, 2003.

\bibitem{Harrison2004}
C.~Harrison, C.~M. Stafford, W.~Zhang, and A.~Karim, ``Sinusoidal phase grating
  created by a tunably buckled surface,'' {\em App. Phys. Lett.}, vol.~85,
  no.~18, pp.~4016--4018, 2004.

\bibitem{Stafford2006}
C.~M. Stafford, B.~D. Vogt, C.~Harrison, D.~Julthongpiput, and R.~Huang,
  ``Elastic moduli of ultrathin amorphous polymer films,'' {\em
  Macromolecules}, vol.~39, no.~15, pp.~5095--5099, 2006.

\bibitem{Brau2011}
F.~Brau, H.~Vandeparre, A.~Sabbah, C.~Poulard, A.~Boudaoud, and P.~Damman,
  ``Multiple-length-scale elastic instability mimics parametric resonance of
  nonlinear oscillators,'' {\em Nature Phys.}, vol.~7, pp.~56--60, 2011.

\bibitem{Huntingdon2013}
M.~D. Huntington, C.~J. Engel, A.~J. Hryn, and T.~W. Odom, ``Polymer
  nanowrinkles with continuously tunable wavelengths,'' {\em ACS Appl. Mater.
  Interfaces.}, vol.~5, no.~13, pp.~6438--6442, 2013.

\bibitem{Bayley2014}
F.~A. Bayley, J.~L. Liao, P.~N. Stavrinou, A.~Chiche, and J.~T. Cabral,
  ``Wavefront kinetics of plasma oxidation of polydimethylsiloxane: limits for
  sub-mm wrinkling,'' {\em Soft Matter}, vol.~10, pp.~1155--1166, 2014.

\bibitem{Wang2015}
Q.~Wang and X.~Zhao, ``A three-dimensional phase diagram of growth-induced
  surface instabilities,'' {\em Sci. Rep.}, vol.~5, p.~8887, 2015.

\bibitem{Leibler1993}
L.~Leibler and K.~Sekimoto, ``On the sorption of gases and liquids in glassy
  polymers,'' {\em Macromolecules}, vol.~26, pp.~6937--6939, 1993.

\bibitem{bar2002shrinkage}
A.~Bar, O.~Ramon, Y.~Cohen, and S.~Mizrahi, ``Shrinkage behaviour of
  hydrophobic hydrogel during dehydration,'' {\em J. Food Eng.}, vol.~55,
  no.~3, pp.~193--199, 2002.

\bibitem{Sekimoto1993}
K.~Sekimoto, ``Temperature hysteresis and morphology of volume phase transition
  of gels,'' {\em Phys. Rev. Lett.}, vol.~70, pp.~4154--4157, 1993.

\bibitem{Gennes2002}
P.-G. de{ }Gennes, ``Solvent evaporation of spin cast films: ``crust''
  effects,'' {\em Eur. Phys. J. E}, vol.~7, pp.~31--34, 2002.

\bibitem{Hennessy2017}
M.~G. Hennessy, G.~L. Ferretti, J.~T. Cabral, and O.~K. Matar, ``A minimal
  model for solvent evaporation and absorption in thin films,'' {\em J.
  Colloid. Interface Sci.}, vol.~488, pp.~61--71, 2017.

\bibitem{talini2023formation}
L.~Talini and F.~Lequeux, ``Formation of glassy skins in drying polymer
  solutions: approximate analytical solutions,'' {\em Soft Matter}, vol.~19,
  no.~30, pp.~5835--5845, 2023.

\bibitem{Okuzono2006}
T.~Okuzono, K.~Ozawa, and M.~Doi, ``Simple model of skin formation caused by
  solvent evaporation in polymer solutions,'' {\em Phys. Rev. Lett.}, vol.~97,
  p.~136103, 2006.

\bibitem{Kodio2017}
O.~Kodio, I.~M. Griffiths, and D.~Vella, ``Lubricated wrinkles: Imposed
  constraints affect the dynamics of wrinkle coarsening,'' {\em Phys. Rev.
  Fluids}, vol.~2, p.~014202, 2017.

\bibitem{Stafford2004}
C.~M. Stafford, C.~Harrison, K.~L. Beers, A.~Karim, E.~J. Amis,
  M.~Vanlandingham, H.-C. Kim, W.~Volksen, R.~D. Miller, and E.~E. Simonyi, ``A
  buckling-based metrologyfor measuring the elastic moduli of polymeric thin
  films,'' {\em Nature Mater.}, vol.~3, pp.~545--550, 2004.

\bibitem{weiss2013creases}
F.~Weiss, S.~Cai, Y.~Hu, M.~Kyoo~Kang, R.~Huang, and Z.~Suo, ``Creases and
  wrinkles on the surface of a swollen gel,'' {\em J. Appl. Phys.}, vol.~114,
  no.~7, 2013.

\bibitem{cerda2003geometry}
E.~Cerda and L.~Mahadevan, ``Geometry and physics of wrinkling,'' {\em Phys.
  Rev. Lett.}, vol.~90, no.~7, p.~074302, 2003.

\bibitem{bowden1998spontaneous}
N.~Bowden, S.~Brittain, A.~G. Evans, J.~W. Hutchinson, and G.~M. Whitesides,
  ``Spontaneous formation of ordered structures in thin films of metals
  supported on an elastomeric polymer,'' {\em Nature}, vol.~393, no.~6681,
  pp.~146--149, 1998.

\bibitem{davis2011mechanics}
C.~S. Davis and A.~J. Crosby, ``Mechanics of wrinkled surface adhesion,'' {\em
  Soft Matter}, vol.~7, no.~11, pp.~5373--5381, 2011.

\bibitem{niven2020emergence}
J.~F. Niven, G.~Chowdhry, J.~S. Sharp, and K.~Dalnoki-Veress, ``The emergence
  of local wrinkling or global buckling in thin freestanding bilayer films,''
  {\em Eur. Phys. J. E}, vol.~43, pp.~1--7, 2020.

\bibitem{genzer2006soft}
J.~Genzer and J.~Groenewold, ``Soft matter with hard skin: From skin wrinkles
  to templating and material characterization,'' {\em Soft Matter}, vol.~2,
  no.~4, pp.~310--323, 2006.

\bibitem{gurmessa2017localization}
B.~J. Gurmessa and A.~B. Croll, ``Localization in an idealized heterogeneous
  elastic sheet,'' {\em Soft matter}, vol.~13, no.~9, pp.~1764--1772, 2017.

\bibitem{Karpitschka2017}
S.~Karpitschka, J.~Eggers, A.~Pandey, and J.~H. Snoeijer, ``Cusp-shaped elastic
  creases and furrows,'' {\em Phys. Rev. Lett.}, vol.~119, p.~198001, 2017.

\bibitem{moore2021nascent}
M.~R. Moore, D.~Vella, and J.~M. Oliver, ``The nascent coffee ring: how solute
  diffusion counters advection,'' {\em J. Fluid Mech.}, vol.~920, p.~A54, 2021.

\end{thebibliography}


\end{document}